\documentstyle[12pt]{article}
\newlength{\mathspace}
\def\b #1{\bar{#1}}
\def\np#1{ Nucl. Phys. B#1}
\def\pr#1    { Phys. Rev. D#1 }
\def\pl#1{ Phys. Lett. B#1}

\def\ijmp#1  { Int. Jour. Mod. Phys. A#1 }
\def\mpl#1   { Mod. Phys. Lett. A#1 }
\def\begineq{\begin{equation}}
\def\endeq{\end{equation}}
\def\eqabegin{\begin{eqnarray}}
\def\eqaend{\end{eqnarray}}
\def\nn{\nonumber}

\begin{document}
\baselineskip=0.7cm
\setlength{\mathspace}{2.5mm}



\begin{titlepage}

    \begin{normalsize}
     \begin{flushright}
  		 SINP-TNP/98-30\\
     \end{flushright}
    \end{normalsize}
    \begin{Large}
       \vspace{.4cm}
       \begin{center}
         {\bf Kaluza-Klein and H-Dyons in String Theory}\\ 
       \end{center}
    \end{Large}

\begin{center}
           
             \vspace{1cm}

            Shibaji R{\sc oy}\footnote{E-mail address: 
            roy@tnp.saha.ernet.in} 

            \vspace{.5cm}

        {\it Saha Institute of Nuclear Physics, 1/AF 
Bidhannagar, Calcutta 700 064, India}\\

      \vspace{2cm}

    \begin{large} ABSTRACT \end{large}
        \par
\end{center}
Kaluza-Klein monopole and H-monopole solutions, which are T-dual to
each other, are the well-known solutions of string theory compactified 
on $T^6$. Since string theory in this case has an S-duality symmetry,
we explicitly construct the corresponding dyonic solutions by expressing
the $D = 4$ string effective action in a manifestly $SL(2, R)$ invariant
form with an $SL(2, R)$ invariant constraint. The Schwarz-Sen charge 
spectrum, the BPS saturated mass formula as well as the stability of these
states are discussed briefly.
\end{titlepage}
\vfil\eject

\begin{large}
\noindent{\bf{1. Introduction}}
\end{large}

\vspace{.5cm}

It is well-known that the five dimensional pure Einstein gravity admits a
solitonic solution known as the Kaluza-Klein (KK) monopole solution first
obtained by Gross, Perry and Sorkin (GPS) [1,2]. Since in this 
construction one
of the spatial coordinates is compactified, this solution can also be
viewed as four dimensional black hole with a magnetic charge. The $U(1)$
gauge field corresponding to the magnetic charge originates in this case
from the isometry of the compact dimension. As the five dimensional pure
gravity is contained as a special case of dimensionally reduced string
theory, obviously, string theory in four/five dimensions also admits KK
monopole solution. String theory in four dimensions admits another kind
of monopole solution known as the H-monopole solution [3,4]. 
The $U(1)$ gauge
field corresponding to the magnetic charge in this case arises from the
dimensional reduction of the second rank antisymmetric tensor $B_{MN}$
contained in the string theory spectrum. These two solutions are in fact
T-dual [5] to each other. In the original ten dimensional theory they
represent the fivebrane solutions compactified on a circle and the
T-duality [6] relates the type IIA (IIB) 
KK monopole to the type IIB (IIA)
Neveu-Schwarz fivebranes. Many interesting dynamical properties of
KK monopoles as well as the world volume theory have been studied in
refs.[7,8,9,10,11,12,13,14]. The world volume theories of 
KK monopole are related to various
supersymmetric gauge theories in (5+1) dimensions and can be used to
study their properties.

Any string theory in four dimensions has been 
conjectured [15] to possess an
exact $SL(2, Z)$ symmetry also known as the S-duality symmetry as a part
of the complete non-perturbative U-duality [16] symmetry. Many evidences 
in favor of this conjecture  have been given in ref.[17]. 
One such evidence
is the prediction of the existence of dyonic solutions corresponding
to both the KK monopole and the H-monopole solutions of string theory.
The proof of existence of these dyonic excitations (in both heterotic
and type II string theory) has been given in refs.[18,19] by arguing that
the degeneracies of the dyonic states match precisely with those of the 
elementary string states. In
this paper, we explicitly construct the dyonic solutions starting 
from the known monopole solutions of type II string theory in a 
simplified setting. We first express the four dimensional string
effective action in a manifestly $SL(2, R)$ invariant form alongwith
an $SL(2, R)$ invariant constraint on the field-strengths. Then we use
this symmetry to rotate the monopole solutions and obtain the
corresponding dyonic solutions. A different kind of dyonic solutions
of both the KK and H-monopoles have been discussed in ref.[20,21],
but the electric and the magnetic charges considered there correspond
to different gauge fields instead of the same gauge field. We also 
obtain the Schwarz-Sen electric-magnetic charge spectrum [22] as well
as the BPS saturated ADM masses for these dyonic solutions. We then 
discuss how the stability of these states can be understood from
the mass formula. Unlike the dyonic solutions discussed in ref.[21],
the dyonic black-hole we obtain has finite Hawking temperature but
zero entropy.

This paper is organized as follows. In section 2, we briefly discuss
the KK monopole solution of GPS and mention how it can be regarded
as a string theory solution in $D = 4$. H-monopole solution is also
discussed in brief. In section 3, we construct the dyonic solutions
by applying $SL(2, R)$ transformation on the monopole solutions. The
Schwarz-Sen charge spectrum and the BPS saturated mass formula are
obtained. We also discuss the stability of the dyonic states. Finally,
our conclusions are presented in section 4.

\vspace{1cm}

\begin{large}
\noindent{\bf{2. KK- and H-Monopole Solutions in String Theory}}
\end{large}

\vspace{.5cm}

KK-monopole was originally obtained by Gross, Perry and Sorkin [1,2] as
a solution of pure Einstein gravity in five dimensions (with one
of the spatial dimensions compactified) where the action has the
form:
\begineq
S_5 = \int\,d^5x \sqrt{-G} R
\endeq
If we denote the compact dimension as $x^4$, then the five dimensional
metric can be decomposed in terms of four dimensional metric as usual
by the following KK ansatz,
\begineq
G_{MN} = \left(\begin{array}{cc} G_{\mu\nu} + A_\mu A_\nu G_{44} &
A_\mu G_{44}\\ A_\nu G_{44} & G_{44}\end{array}\right)
\endeq
where $M,\,N = 0,\,1,\ldots,4$ and $\mu,\,\nu = 0,\,1,\,2,\,3$. Also
all the four dimensional fields here are independent of $x^4$.
Now the five dimensional action (1) reduces to the four-dimensional
action as follows:
\begineq
S_4 = \int\, d^4x \sqrt{-{\b G}}\left[R - \frac{1}{2} \partial_\mu
{\tilde \phi} \partial^\mu {\tilde \phi} - \frac{1}{4} e^{-\sqrt{3}
{\tilde \phi}} F_{\mu\nu} F^{\mu\nu}\right]
\endeq
where the metric ${\b G}_{\mu\nu} = G^{1/2}_{44} G_{\mu\nu}$, the 
scalar ${\tilde \phi} = - \frac{\sqrt{3}}{2}\log G_{44}$ and $F_{
\mu\nu} = \partial_\mu A_\nu - \partial_\nu A_\mu$.

The equations of motion obtained from action (3) has a magnetically 
charged black hole solution which in the extremal limit takes the
form [23,24,25]:
\eqabegin
ds^2 &=& -\left(1 + \frac{P}{r}\right)^{-1/2} dt^2 + \left(1 + \frac{P}
{r}\right)^{1/2}\left[dr^2 + r^2 d\Omega_2^2\right]\nn\\
e^{2{\tilde \phi}} &=& \left(1 + \frac{P}{r}\right)^{\sqrt{3}} \qquad
{\rm or} \qquad e^{\frac{2{\tilde \phi}}{\sqrt{3}}}\,\,\,=\,\,\,
\left(1 + \frac{P}{r}\right)\,\,\,=\,\,\, G_{44}^{-1}
\eqaend
Here $ds^2$ is written in terms of the canonical metric $\b {G}_{\mu\nu}$
and in terms of $G_{\mu\nu}$ it is given by
\eqabegin
ds^2 &=& - dt^2 + \left(1 + \frac{P}{r}\right)\left[dr^2 + r^2 d\Omega_2^2
\right]\nn\\
&=& - dt^2 + G_{44}^{-1} \left[dr^2 + r^2 d\Omega_2^2\right]
\eqaend
where $d\Omega_2^2 = d\theta^2 + \sin^2\theta d\varphi^2$ and $P$ is the
magnetic charge of the black hole. The solution here is written in terms of
spherical polar coordinates $r$, $\theta$, $\varphi$ denoted as 1, 2, 3
respectively. Thus $F_{23} = P \sin\theta$. This is the KK monopole solution
of GPS in four dimensions. Since the $U(1)$ gauge field in this case has
the form $A_3 = P(1 - \cos\theta)$, the above solution can be written
in terms of Taub-NUT metric in five dimensions as in [2]. 
Let us now try to understand how the above
solution arises in string theory. The low energy effective action of any
string theory in $D = 5$ has the following  form in common:
\begineq
S_5^{(st)} = \int\,d^5x \sqrt{-G} e^{-2\Phi}\left[R + 4 \partial_M
\Phi\partial^M \Phi - \frac{1}{12} H_{MNP}H^{MNP}\right]
\endeq
Here $G_{MN}$, $\Phi$ and $B_{MN}$ are respectively the five dimensional
metric, dilaton and Kalb-Ramond antisymmetric tensor field, with the
field strength $H_{MNP} = \partial_M B_{NP} + \partial_N B_{PM} + 
\partial_P B_{MN}$. The rest of the fields which arise from the dimensional
reduction are set to zero. If we now further set $H_{MNP} = 0$, then 
with the same KK ansatz (2) of the metric, we can write the reduced action
in the form:
\begineq
S_4^{(st)} = \int\,d^4x \sqrt{-G} e^{-2\phi}\left[R + 4\partial_\mu\phi
\partial^\mu\phi - \partial_\mu\sigma \partial^\mu\sigma - \frac{1}{4}
e^{-2\sigma} F_{\mu\nu}^{(1)} F^{(1)\,\mu\nu}\right]
\endeq
where the four dimensional dilaton $\phi = \Phi + \frac{1}{2}\sigma$ and
the scalar $\sigma$ is given by $G_{44} = e^{-2\sigma}$. We have also
renamed the gauge field $A_\mu$ as $A_\mu^{(1)}$ for later convenience.
Thus we notice that the four dimensional string action contains two scalars
$\phi$ and $\sigma$ instead of one, ${\tilde \phi}$, as in pure gravity case
(3). The solution of the equations of motion following from (7) is also given
in terms of two scalars as follows [25]:
\eqabegin
ds^2 &=& - dt^2 + e^{2\phi + \sigma}\left[dr^2 + r^2 d\Omega_2^2\right]\nn\\
e^{2\phi} &=& e^\sigma\,\,\,=\,\,\, \left(1 + \frac{P^{(1)}}{r}\right)^
{1/2}
\eqaend
where we have renamed the KK monopole charge $P$ as $P^{(1)}$. Note that 
the solution $ds^2$ is written in terms of  string metric $G_{\mu\nu}$.
In terms of canonical metric $\b {G}_{\mu\nu} = e^{-2\phi} G_{\mu\nu}$,
the solution reduces to
\begineq
ds^2 = - e^{-2\phi} dt^2 + e^\sigma \left[dr^2 + r^2 d\Omega_2^2\right]
\endeq
We find that the solution (8) and (9) are precisely identical with
(5) and (4) since $\sigma = \frac{1}{\sqrt {3}} {\tilde \phi}$. Note
further that since $\phi = \frac{1}{2} \sigma$ is the solution, the
five dimensional dilaton $\Phi$ is trivial (but the four dimensional
dilaton is not) as expected. Thus we note that although we started out
with different actions namely, (3) and (7), we end up with the same KK
monopole solutions and this clarifies how KK monopole arises as a 
solution in string theory. We now discuss the H-monopole solution in
string theory. Note from (6) that if instead of setting $H_{MNP}$ to
zero, we keep the component $B_{\mu4} = A_\mu ^{(2)}$, the reduced action
would take the form with $A_\mu^{(1)} = 0$, as,
\begineq
{\tilde S}_4^{(st)} = \int\, d^4x \sqrt {-G} e^{-2\phi} \left[R +
4\partial_\mu\phi \partial^\mu\phi - \partial_\mu \sigma \partial^\mu
\sigma - \frac{1}{4} e^{2\sigma} F_{\mu\nu}^{(2)} F^{(2)\,\mu\nu}\right]
\endeq
The equations of motion following from (10) has the solution,
\eqabegin
ds^2 &=& - dt^2 + e^{2\phi - \sigma}\left[dr^2 + r^2 d\Omega_2^2\right]\nn\\
e^{2\phi} &=& e^{-\sigma} \,\,\,=\,\,\, \left(1 + \frac{P^{(2)}}{r}
\right)^{1/2}
\eqaend
Here $P^{(2)}$ is the magnetic charge associated with H-monopole i.e.
$F_{23}^{(2)} = P^{(2)} \sin\theta$. In the canonical metric the
solution can be written as,
\begineq
ds^2 = - e^{-2\phi} dt^2 + e^{-\sigma}\left[dr^2 + r^2 d\Omega_2^2\right]
\endeq
Note that since here $2\phi = -\sigma$ is the solution, the five
dimensional dilaton $\Phi$ is not trivial and therefore this solution
is strictly a string theory solution which can not be obtained from
pure gravity. This can also be understood since the gauge field 
$A_\mu^{(2)}$ originates in this case from the dimensional reduction
of $B_{MN}$ which is a field contained only in string theory spectrum.
It should be mentioned here that the solution (11) can be obtained from 
(8)
 by the following transformations,
\begineq
G_{\mu\nu} \rightarrow G_{\mu\nu}, \qquad
\sigma \rightarrow -\sigma, \qquad \phi \rightarrow \phi,
\qquad A_\mu^{(1)} \rightarrow A_\mu^{(2)}, \qquad P^{(1)} \rightarrow
P^{(2)}
\endeq
Although neither the actions (7) nor (10) possess this symmetry, the
full string theory action indeed has this symmetry, the T-duality
symmetry (which is $O(1,1)$ symmetry in this case). Thus the string 
theory admits both the KK and H-monopole solutions which are related
to each other by the T-duality transformations. 

\vspace{1cm}

\begin{large}
\noindent{\bf{3. Dyonic Solutions}}
\end{large}

\vspace{.5cm}

Starting from the monopole solutions discussed in the previous section,
we, in this section, will construct the corresponding dyonic solutions
having both magnetic and electric charges. In order to obtain these
solutions we will use the $SL(2, R)$ symmetry of the four dimensional
string effective action. The relevant four dimensional string action
containing both the gauge fields $A_\mu^{(1)}$ and $A_\mu^{(2)}$ as
well as the antisymmetric tensor field $B_{\mu\nu}$ has the form
\eqabegin
S^{(st)} &=& \int\, d^4x \sqrt{-G} e^{-2\phi}\left[R + 4 \partial_\mu
\phi \partial^\mu\phi - \partial_\mu\sigma \partial^\mu\sigma -
\frac{1}{4} e^{-2\sigma} F_{\mu\nu}^{(1)} F^{(1)\,\mu\nu}\right.\nn\\
& &\qquad\qquad\qquad\qquad\left. -\frac{1}{4} e^{2\sigma} F_{\mu\nu}^{(2)}
F^{(2)\,\mu\nu} - \frac{1}{12} H_{\mu\nu\lambda} H^{\mu\nu\lambda}
\right]
\eqaend
Note that when both the gauge fields $A_\mu^{(1)}$ and $A_\mu^{(2)}$
are non-zero, the reduced form of the field strength $H_{\mu\nu\lambda}$
is given as\footnote[1]{The antisymmetric tensor field $B_{\mu\nu}$ here
is a modified form of the dimensionally reduced $B_{\mu\nu}$ [26].},
\begineq
H_{\mu\nu\lambda} = \partial_\mu B_{\nu\lambda} - \frac{1}{2}
\left(A_\mu^{(1)} F_{\nu\lambda}^{(2)} + A_\mu^{(2)} F_{\nu\lambda}^{(1)}
\right) + {\rm cyc.\,\,in\,\,}\mu\nu\lambda
\endeq
So, the action (14) is invariant under the T-duality transformation (13)
alongwith
\begineq
A_{\mu}^{(2)} \rightarrow A_\mu^{(1)}, \qquad B_{\mu\nu} \rightarrow
B_{\mu\nu}, \qquad P^{(2)} \rightarrow P^{(1)}
\endeq
Note that $H^2$ term is invariant under the transformations (13) and (16)
by itself.
The transformations (13) and (16) constitute the complete T-duality
or $O(1, 1)$  symmetry of the theory. When $H_{\mu\nu\lambda} = 0$ the 
equations of motion\footnote[2]{Here the equations of motion are
obtained after rewriting the action in the Einstein metric, where
Einstein metric $\b {G}_{\mu\nu} = e^{-2\phi} G_{\mu\nu}$.} following 
from (14) are as given below:
\eqabegin
& & \nabla^2 \phi + \frac{1}{8} e^{-2\phi}\left[e^{-2\sigma} (F^{(1)})^2
+ e^{2\sigma} (F^{(2)})^2\right]\,\,\, =\,\,\, 0\\
& & \nabla^2 \sigma + \frac{1}{4} e^{-2\phi}\left[e^{-2\sigma} (F^{(1)})^2
- e^{2\sigma} (F^{(2)})^2\right]\,\,\, =\,\,\, 0\\
& & \nabla_\mu\left(e^{-2\phi - 2\sigma} F^{(1)\,\mu\nu}\right)\,\,\, 
=\,\,\, 0\\
& & \nabla_\mu\left(e^{-2\phi + 2\sigma} F^{(2)\,\mu\nu}\right)\,\,\, 
=\,\,\, 0\\
& & R_{\mu\nu}\,\,\,=\,\,\, 2\partial_\mu\phi\partial_\nu\phi + 
\partial_\mu \sigma \partial_\nu\sigma + \frac{1}{2} e^{-2\phi}\left(
e^{-2\sigma} F_{\mu\rho}^{(1)} F_\nu^{(1)\,\,\,\rho} + e^{2\sigma}
F_{\mu\rho}^{(2)} F_\nu^{(2)\,\,\,\rho}\right)\nn\\
& &\qquad\qquad\qquad -\frac{1}{8} \b {G}_{\mu\nu} e^{-2\phi}\left(
e^{-2\sigma} (F^{(1)})^2 + e^{2\sigma} (F^{(2)})^2\right)
\eqaend
The equations of motion (17)--(21) can be easily solved by assuming
the form of the metric to be static, spherically symmetric which becomes
flat asymptotically. In fact, the supersymmetric BPS saturated solution
of the above equations of motion has already been obtained by Cvetic
and Youm [21] which has the following form:
\eqabegin
ds^2 &=& - \left(f_1 f_2\right)^{-1/2} dt^2 + \left(f_1 f_2\right)^{1/2}
\left[dr^2 + r^2 d\Omega_2^2\right]\nn\\
e^{2\phi} &=& \left(f_1 f_2\right)^{1/2}; \qquad e^{-\sigma}\,\,\,=\,\,\,
\left(\frac{f_2}{f_1}\right)^{1/2}
\eqaend
where $f_i = \left(1 + \frac{P^{(i)}}{r}\right)$ with $i = 1,\,2$. Here
$P^{(i)}$'s are the magnetic charges associated with the gauge fields
$A_\mu^{(i)}$. The solution is indeed invariant under the T-duality
transformations (13) and (16). Note from (22) that $P^{(2)} = 0$
corresponds to KK monopole solution (8) whereas $P^{(1)} = 0$ corresponds
to H-monopole solution (11) as discussed in the previous section.

We would like to point out that for $H_{\mu\nu\lambda} \neq 0$ the 
action (14) as well as the 
equations of motion following from it has a larger symmetry
than what has already been noted in (13) and (16). In fact, apart
from the T-duality symmetry, it also has an S-duality  symmetry and
we will use this symmetry to obtain the dyonic solutions.

We first note that the action (14) can be written in an $O(1, 1)$
invariant form [27] as follows:
\eqabegin
S^{(st)} &=& \int\, d^4x \sqrt{-G} e^{-2\phi}\left[R + 4\partial_\mu
\phi \partial^\mu\phi + \frac{1}{8} {\rm tr}\,\partial_\mu M 
\partial^\mu M^{-1}\right.\nn\\
& & \left. \qquad\qquad\qquad\qquad -\frac{1}{4} {\cal F}_{\mu\nu}^T
M^{-1} {\cal F}^{\mu\nu} - \frac{1}{12} H_{\mu\nu\lambda}
H^{\mu\nu\lambda}\right]
\eqaend
where $M = \left(\begin{array}{cc} e^{2\sigma} & 0\\ 0 & e^{-2\sigma}
\end{array}\right)$ is an $O(1, 1)$ matrix satisfying $M^T \eta M =
\eta$, with $\eta = \left(\begin{array}{cc} 0 & 1\\ 1 & 0
\end{array}\right)$ and ${\cal F}_{\mu\nu} = \left(\begin{array}{c}
F_{\mu\nu}^{(1)}\\ F_{\mu\nu}^{(2)}\end{array}\right)$. The action
(23) is invariant under a global $O(1, 1)$ transformation
\eqabegin
M &\rightarrow& \Omega M \Omega^T, \qquad {\cal F}_{\mu\nu}\,\,\
\rightarrow\,\,\ \Omega {\cal F}_{\mu\nu}, \qquad G_{\mu\nu}\,\,\,
\rightarrow\,\,\, G_{\mu\nu},\nn\\
\phi & \rightarrow & \phi \qquad {\rm and} \qquad B_{\mu\nu} \,\,\,
\rightarrow\,\,\, B_{\mu\nu}
\eqaend
where $\Omega$ is an $O(1, 1)$ matrix.
Note that in this particular case $\Omega = \eta$. By writing (23)
in Einstein metric,
\eqabegin
\b {S}^{(st)} &=& \int\, d^4x \sqrt{-\b {G}}\left[R - 2\partial_\mu
\phi \partial^\mu\phi + \frac{1}{8} {\rm tr}\,\partial_\mu M 
\partial^\mu M^{-1}\right.\nn\\
& &\qquad\qquad\qquad\qquad \left. -\frac{1}{4} e^{-2\phi} 
{\cal F}_{\mu\nu}^T
M^{-1} {\cal F}^{\mu\nu} - \frac{1}{12} e^{-4\phi} H_{\mu\nu\lambda}
H^{\mu\nu\lambda}\right]
\eqaend
we find that the equations of motion derived from (25) can also be 
obtained from an alternative action [28,17]:
\eqabegin
{S}^{(alt)} &=& \int\, d^4x \sqrt{-\b {G}}\left[R - 
\frac{1}{2\lambda_2^2} \partial_\mu\lambda 
\partial^\mu\b {\lambda} + \frac{1}{8} {\rm tr}\,\partial_\mu M 
\partial^\mu M^{-1}\right.\nn\\
& &\qquad\qquad\qquad\qquad \left. -\frac{1}{4} \lambda_2 
{\cal F}_{\mu\nu}^T
M^{-1} {\cal F}^{\mu\nu} - \frac{1}{4} \lambda_1 {\cal F}_{\mu\nu}^T
\eta {\tilde {\cal F}}^{\mu\nu}\right]
\eqaend
where we have defined
\begineq
H^{\mu\nu\lambda} = - \frac{1}{\sqrt {-\b {G}}} e^{4\phi} 
\epsilon^{\mu\nu\lambda\rho}\partial_\rho a
\endeq
with `$a$' a pseudoscalar called axion. (27) in fact follows from the
equation of motion of $H_{\mu\nu\lambda}$ in (25). Also, $\lambda$ is a 
complex scalar defined as,
\begineq
\lambda = a + i e^{-2\phi} = \lambda_1 + i \lambda_2
\endeq
and ${\tilde {\cal F}}_{\mu\nu} = \frac{1}{2} \sqrt{-\b{G}} 
\epsilon_{\mu\nu\lambda\rho} {\cal F}^{\lambda\rho}$. The equations
of motion following from (26) are:
\eqabegin
& & \frac{1}{\lambda_2^2}\nabla_\mu\partial^\mu\lambda + \frac{i}
{\lambda_2^3}\partial_\mu\lambda\partial^\mu\lambda - \frac{1}{4}
{\cal F}_{\mu\nu}^T \eta {\tilde {\cal F}}^{\mu\nu} - \frac{i}{4}
{\cal F}_{\mu\nu}^T M^{-1} {\cal F}^{\mu\nu}\,\,\,=\,\,\,0\\
& & R_{\mu\nu}\,\,\,=\,\,\,\frac{1}{4\lambda_2^2}\left(\partial_\mu
\lambda\partial_\nu\b{\lambda} + \partial_\nu\lambda\partial_\mu
\b{\lambda}\right) + \frac{1}{2} \lambda_2 {\cal F}_{\mu\rho}^T M^{-1}
{\cal F}_\nu^{\,\,\,\rho}\nn\\
& & \qquad\qquad\qquad -\frac{1}{8} \lambda_2 \b{G}_{\mu\nu}{\cal F}_
{\rho\sigma}^T M^{-1} {\cal F}^{\rho\sigma} - \frac{1}{8} {\rm tr}
\partial_\mu M \partial_\nu M^{-1}\\
& & \nabla_\mu\left(\lambda_2 M^{-1} {\cal F}^{\mu\nu} + \lambda_1
\eta {\tilde {\cal F}}^{\mu\nu}\right) \,\,\,=\,\,\,0
\eqaend
and the $\sigma$ equation is as given in (18). It can now be 
checked that the above equations of motion are invariant under the
global $SL(2, R)$ transformations as given below [29,28,17]:
\eqabegin
\lambda &\rightarrow& \frac{a\lambda + b}{c\lambda + d},\qquad
\b{G}_{\mu\nu}\,\,\,\rightarrow \,\,\, \b{G}_{\mu\nu}, \qquad
M\,\,\,\rightarrow\,\,\,M\nn\\
{\cal F}_{\mu\nu} &\rightarrow& (c\lambda_1 + d) {\cal F}_{\mu\nu}
- c\lambda_2 M \eta {\tilde {\cal F}}_{\mu\nu}
\eqaend
where the global $SL(2, R)$ transformation matrix $\Lambda = \left(
\begin{array}{cc} a & b\\ c & d\end{array}\right)$, with $ ad - bc = 1$.
However, we note that the action (26) is not $SL(2, R)$ invariant. Let
us next rewrite the action in an $SL(2, R)$ invariant [16,30] fashion
\footnote[2]{An $SL(2, R)$ invariant four dimensional string effective
action has also been constructed in [31] using a different method.}. 
In order 
to do this we define
\begineq
{\cal M} = \left(\begin{array}{cc} e^{-2\phi} + a^2 e^{2\phi} &
a e^{2\phi} \\ a e^{2\phi} & e^{2\phi}\end{array}\right)
\endeq
and
\begineq
{\cal H}_{\mu\nu} = \left(\begin{array}{c}{\cal F}_{\mu\nu} \\
{\cal G}_{\mu\nu}\end{array}\right)
\endeq
where
\begineq
{\cal G}_{\mu\nu} = {\hat {\cal F}}_{\mu\nu} - a {\cal F}_{\mu\nu}
\endeq
Here ${\hat {\cal F}}_{\mu\nu}$ is a pair of new fields we introduce 
which will be
related to the known fields through a constraint as we will see below.
The action (26) can now be written in a manifestly $SL(2, R)$ invariant 
form as follows,
\begineq
S = \int\, d^4x \sqrt{-\b{G}}\left[R + \frac{1}{8}{\rm tr}\,\partial_\mu
M \partial^\mu M^{-1} + \frac{1}{4} {\rm tr}\,\partial_\mu {\cal M}
\partial^\mu {\cal M}^{-1} - \frac{1}{4} {\cal H}_{\mu\nu}^T {\cal M}
M^{-1} {\cal H}^{\mu\nu}\right]
\endeq
where the $SL(2, R)$ transformations are given as,
\begineq
\b{G}_{\mu\nu} \rightarrow \b{G}_{\mu\nu}, \qquad M \rightarrow M,
\qquad {\cal M} \rightarrow \Lambda {\cal M} \Lambda^T, \qquad
{\cal H}_{\mu\nu} \rightarrow (\Lambda^T)^{-1} {\cal H}_{\mu\nu}
\endeq
where $\Lambda$ is an $SL(2, R)$ matrix.
In order to see the equivalence between (26) and (36), we note that the
Bianchi identities and the equations of motion of the field strengths
and the gauge fields following from (36) are: 
\eqabegin
\nabla_\mu {\tilde {\cal F}}^{\mu\nu} &=& 0\\
\nabla_\mu {\tilde {\cal G}}^{\mu\nu} &=& 0 \quad \Rightarrow
\nabla_\mu\left({\tilde {\hat {\cal F}^{\mu\nu}}} 
- a {\tilde {\cal F}}^{
\mu\nu}\right)\,\,\,=\,\,\,0
\eqaend
and
\eqabegin
\nabla_\mu\left(e^{2\phi} M^{-1} {\hat {\cal F}}^{\mu\nu}\right) &=& 0
\\
\nabla_\mu\left(e^{-2\phi} M^{-1} {\cal F}^{\mu\nu} + a e^{2\phi}
M^{-1} {\hat {\cal F}}^{\mu\nu}\right) &=& 0
\eqaend
Now it can be checked that if we impose the constraint
\begineq
{\hat {\cal F}}^{\mu\nu} = e^{-2\phi} M \eta {\tilde {\cal F}}^{\mu\nu}
\endeq
then (39) and (41) reduce to the equations of motion (31) derived 
from (26), whereas, (40) reduces to the Bianchi identity (38). All other
equations of motion can also be shown to remain unaffected. Thus, we
conclude that the action (26) is equivalent to action (36) subject
to the constraint (42). Now the constraint (42) can also be written
in an $SL(2, R)$ invariant form as,
\begineq
{\tilde {\cal H}}_{\mu\nu} = M \eta \Sigma {\cal M} {\cal H}_{\mu\nu}
\endeq
where $\Sigma = \left(\begin{array}{cc} 0 & 1\\ -1 & 0\end{array}\right)$
is the $SL(2, R)$ metric satisfying $\Lambda \Sigma \Lambda^T =
\Lambda^T \Sigma \Lambda = \Sigma$. We would like to point out that
the action (36) contains an $SL(2, R)$ doublet of field strengths
${\cal F}_{\mu\nu}$ and ${\cal G}_{\mu\nu}$, which makes it manifestly
$SL(2, R)$ invariant. On the other hand, action (26) does not contain
the $SL(2, R)$ doublet of field strengths, but the equations of motion
following from it contains the $SL(2, R)$ doublet ${\cal F}_{\mu\nu}$
and $e^{-2\phi} M \eta {\tilde {\cal F}}_{\mu\nu}$, which can be seen 
from (42). That is why the equations of motion of (26) are $SL(2, R)$ 
invariant eventhough the action itself is not.
 
The $SL(2, R)$ transformation matrix which takes the asymptotic value 
(the subscript `0' will always denote the asymptotic value) of
the complex moduli $\lambda_0 = i$ (corresponding to $a_0 = \phi_0 = 0$)
to an arbitrary value $\lambda_0$ has the form [32, 33]
\begineq
\Lambda = \left(\begin{array}{cc} e^{-\phi_0} \cos\alpha + a_0 \sin\alpha
e^{\phi_0} & -\sin\alpha e^{-\phi_0} + a_0 \cos\alpha e^{\phi_0}\\
e^{\phi_0} \sin\alpha & e^{\phi_0} \cos\alpha\end{array}\right)
\endeq
Here `$\alpha$' is an arbitrary parameter and will be fixed later.
The transformation of the field strengths from (37) then is given as,
\eqabegin
{\cal F}_{\mu\nu} &\rightarrow& e^{\phi_0} \cos\alpha {\cal F}_{\mu\nu}
- e^{\phi_0} \sin\alpha {\cal G}_{\mu\nu}\\
{\cal G}_{\mu\nu} &\rightarrow& \left(e^{-\phi_0} \sin\alpha - a_0
e^{\phi_0} \cos\alpha\right){\cal F}_{\mu\nu} + \left(e^{-\phi_0}
\cos\alpha + a_0 e^{\phi_0} \sin\alpha\right) {\cal G}_{\mu\nu}
\eqaend
Note that the initial configuration i.e. the monopole solution (22)
has only ${\cal F}_{\mu\nu}$ where $F_{23}^{(i)} = P^{(i)} \sin\theta$,
for $i = 1, 2$. So, the magnetic charges $P^{(1)}$, $P^{(2)}$ are 
integers measured in some basic units. In other words, $P^{(1)} = mP$
and $P^{(2)}= nP$, where $m , n$ are integers and $P$ is the charge unit
which is set to 1 from now on. Now as we make the $SL(2, R)$
transformations (45), (46), they will no longer remain integers. So, 
we modify $P^{(1)}$ and $P^{(2)}$ by $\Delta_1^{1/2}$ and 
$\Delta_2^{1/2}$ and demand that the charges remain integers after
the transformation. $\Delta_1$ and $\Delta_2$ will be fixed soon. 
Thus from (45) and (46) we obtain,
\eqabegin
\cos\alpha &=& e^{-\phi_0} \Delta_1^{-1/2} p_1 \,\,\,=\,\,\, e^{-\phi_0}
\Delta_2^{-1/2} p_2\\
\sin\alpha &=& \left(q_1 + a_0 p_1\right) e^{\phi_0} \Delta_1^{-1/2}
\,\,\,=\,\,\, \left(q_2 + a_0 p_2\right) e^{\phi_0} \Delta_2^{-1/2}
\eqaend
where $(p_i, q_i)$ are integers measuring the number of units of 
magnetic and `electric' charges of the dyonic solution. Note that the
`electric' charge mentioned here is an auxiliary `electric' charge
corresponding to ${\cal G}_{\mu\nu}$. The true electric charge of the
theory is given later in eq.(59) which is non-integral. Thus the 
dyonic solution is characterized by two pairs of integers corresponding
to the magnetic and `electric' charges associated with ${\cal F}_{\mu\nu}$
and ${\cal G}_{\mu\nu}$.
(47) and (48) determines the value of $\Delta_1$ and $\Delta_2$ as,
\eqabegin
\Delta_i &=& e^{-2\phi_0} p_i^2 + e^{2\phi_0} \left(q_i + 
a_0 p_i\right)^2\nn\\
&=& (p_i,\,q_i) {\cal M}_0 \left(\begin{array}{c} p_i\\ q_i\end{array}
\right)
\eqaend
As is clear, `$i$' is not summed over in the r.h.s. of eq.(49).
Since the charges  transform as $\left(\begin{array}{c} p_i \\ q_i 
\end{array}\right) \rightarrow
(\Lambda^T)^{-1} \left(\begin{array}{c} p_i \\ q_i
\end{array}\right)$, we find that (49) is $SL(2, Z)$ invariant. Note
from (49) that in order to maintain the charge vector to be integer
valued the $SL(2, R)$ transformation would have to be restricted to
$SL(2, Z)$ i.e. integer valued, but then the asymptotic value of the
dilaton ($\phi$) and the axion ($a$) can not be maintained to a fixed
value. We here construct the dyonic solution for a fixed but arbitrary
asymptotic value of the background fields. The transformed value of 
the complex moduli and the field strengths
are given as
\eqabegin
\lambda' &=& \frac{a_0 \Delta_1 A + p_1 q_1 e^{-2\phi_0} (A - 1) +
i \Delta_1 A^{1/2} e^{-2\phi_0}}{p_1^2 e^{-2\phi_0} + A e^{2\phi_0}
(a_0 p_1 + q_1)^2}\\
&=& \frac{a_0 \Delta_2 A + p_2 q_2 e^{-2\phi_0} (A - 1) +
i \Delta_2 A^{1/2} e^{-2\phi_0}}{p_2^2 e^{-2\phi_0} + A e^{2\phi_0}
(a_0 p_2 + q_2)^2}
\eqaend
and
\eqabegin
{\cal F'}_{23} &=& \left(\begin{array}{c} p_1  \sin\theta \\
p_2  \sin\theta\end{array}\right)\\
{\cal G'}_{23} &=& \left(\begin{array}{c} q_1  \sin\theta\\
q_2  \sin\theta\end{array}\right)
\eqaend
In (50) and (51) the function $A$ is defined as,
\begineq
A = \left[\left(1 + \frac{\Delta_1 ^{1/2} }{r}\right)\left(1 +
\frac{\Delta_2^{1/2} }{r}\right)\right]^{-1/2}
\endeq
Note that asymptotically $A \rightarrow 1$ and therefore, from (50)
and (51) we have $\lambda \rightarrow \lambda_0$ as expected. Finally,
the canonical metric for the dyonic solution is given as,
\eqabegin
ds^2 &=& -\left[\left(1 + \frac{\Delta_1 ^{1/2} 
}{r}\right)\left(1 +
\frac{\Delta_2^{1/2} }{r}\right)\right]^{-1/2} dt^2\nn\\ 
& & \qquad\qquad\qquad +\left[\left(1 + \frac{\Delta_1 ^{1/2} 
}{r}\right)\left(1 +
\frac{\Delta_2^{1/2} }{r}\right)\right]^{1/2} \left[dr^2 + 
r^2 d\Omega_2^2\right]
\eqaend
and
\begineq
e^{-\sigma} = \left[\frac{ 1 + \frac{\Delta_2^{1/2} }{r}}
{1 + \frac{\Delta _1^{1/2} }{r}}\right]^{1/2}
\endeq
Thus starting from the magnetic monopole solution (22) we have obtained
the dyonic solution carrying both magnetic and electric charges given by
the field configurations (50)--(56). The magnetic charge 
$\Delta_2^{1/2} = 0$
corresponds to the KK dyonic solution whereas 
$\Delta_1^{1/2} = 0$ corresponds
to the H dyonic solution. The BPS saturated ADM mass of the dyonic 
solution can be easily calculated [34] from (55) which is given by,
\begineq
M^2 = \frac{1}{16}\left(\Delta_1^{1/2}  + \Delta_2^{1/2} 
\right)^2
\endeq
We note from above that
the mass formula in (57)
can be written in a manifestly $SL(2, Z)$ and $O(1, 1, Z)$ invariant form
as [17,22] 
\begineq
M^2 = \frac{1}{16} (p_i,\, q_i) {\cal M}_0 \left(M_0 + \eta\right)_{ij}
\left(\begin{array}{c} p_j \\ q_j\end{array}\right)
\endeq
$M_0$ in (58) is actually an identity matrix since $e^\sigma \rightarrow
1$ asymptotically as can be seen from (56). This is the Schwarz-Sen mass 
formula in this case. The electric, magnetic charge spectrum for the 
dyonic solutions can also be obtained from (35) and (42) as,
\begineq
\left(Q_i^{(mag)},\, Q_i^{(el)}\right) = \left(p_i,\, e^{2\phi_0} (M_0
\eta)_{ij}\left(q_j + a_0 p_j\right)\right)
\endeq
This charge spectrum has been obtained by Schwarz and Sen in ref.[22] and
they satisfy the Dirac-Schwinger-Zwanziger-Witten quantization rule [35].

Now in order to understand the stability [36,17] of the 
dyonic solution, we
first note from (58) that the masses of such solution are characterized
by two pairs of integers $(p_1,\,q_1,\,p_2,\,q_2)$ and 
they satisfy as usual
for a BPS state
the triangle inequality of the form,
\begineq
M_{(p_1,q_1,p_2,q_2)} + M_{(p_1',q_1',p_2',q_2')} \geq
M_{(p_1+p_1',q_1+q_1',p_2+p_2',q_2+q_2')}
\endeq
The equality holds when
\begineq
\left(p_1 + p_2\right)\left(q_1' + q_2'\right) = \left(p_1' + p_2'\right)
\left(q_1 + q_2\right)
\endeq
Hence we notice that when sum of the magnetic charges $\left(p_1 
+ p_2\right)$ is 
relatively  prime to the sum of the `electric' charges 
$\left(q_1 + q_2\right)$,
the dyonic state will be stable since it will be prevented from
decaying into lower mass state by the inequality (60).

Finally, we note from (58) that unlike the dyonic solution 
considered in ref.[21], the solution considered here has zero area of 
the event horizon i.e. zero entropy but finite Hawking temperature
\begineq
T_H = \frac{1}{4\pi \sqrt{\Delta_1^{1/2}  \Delta_2^{1/2} }}
\endeq

\vspace{1cm}

\begin{large}
\noindent{\bf{4. Conclusion}}
\end{large}

\vspace{.5cm}

To summarize, we have briefly discussed the KK magnetic monopole solution
in five dimensional pure Einstein gravity with one of the spatial
dimensions compactified and then showed how this solution
arises in string theory. We have also discussed another kind of monopole
solution known as H-monopole solution in string theory. These two solutions
are related to each other by a T-duality transformation in string theory.
Next, we considered the magnetic monopole solution when both KK gauge field
and the gauge field originating from the dimensional reduction of Kalb-Ramond
antisymmetric tensor field are present. Then by including the $H^2$ term we
have shown that the full string theory effective action can be expressed in
a manifestly $SL(2, R)$ invariant form with an $SL(2, R)$ invariant constraint.
By using this symmetry we have explicitly constructed the corresponding 
dyonic solution. We have obtained the BPS saturated ADM mass formula and the
Schwarz-Sen electric-magnetic charge spectrum for this solution. The dyonic
solution is characterized by two pairs of integers. By using the mass formula
we have shown that when sum of the magnetic charges 
is relatively prime to the sum of the `electric' charges 
the dyonic solution is stable. The stability can be 
understood from a triangle inequality relation satisfied by the masses of the
dyonic states. We have mentioned that unlike the dyonic solution considered
in ref.[21], the solution we described has zero entropy, but finite Hawking
temparature.

\vspace{1cm}

\begin{large}
\noindent{\bf{Acknowledgements}}
\end{large}

\vspace{.5cm}

I would like to thank Ashoke Sen for very useful discussions. I would also
like to thank J. X. Lu for some interesting comments. I am grateful to the
referee for the comments which has helped me to improve the paper.

\vspace{1cm}

\begin{large}
\noindent{\bf {References}}
\end{large}

\vspace{.5cm}

\begin{enumerate}
\item R. Sorkin, Phys. Rev. Lett. 51 (1983) 87.
\item D. Gross and M. Perry, \np 226 (1983) 29.
\item R. Khuri, \pl 259 (1991) 261; \pl 294 (1992) 325; \np 387 (1992) 315.
\item J. Gauntlett, J. Harvey and J. Liu, \np 409 (1993) 363.
\item T. Banks, M. Dine, H. Dijkstra and W. Fischler, \pl 212 (1988) 45.
\item A. Giveon, M. Porrati and E. Rabinovici, Phys. Rep. C244 (1994) 77;
E. Alvarez, L. Alvarez-Gaume and Y. Lozano, {\it An introduction to T-duality
in string theory}, hep-th/9410237.
\item P. Ruback, Comm. Math. Phys. 107 (1986) 93.
\item C. Hull, {\it Gravitational duality, branes and charges}, hep-th/9705162.
\item E. Bergshoeff, B. Janssen and T. Ortin, \pl 410 (1997) 132.
\item Y. Imamura, \pl 414 (1997) 242.
\item A. Sen, Adv. Theor. Math. Phys. 1 (1998) 115.
\item A. Hanany and G. Lifschytz, {\it M(atrix) theory on $T^6$ and a m(atrix)
theory description of KK monopoles}, hep-th/9708037.
\item R. Gregory, J. Harvey and G. Moore, {\it Unwinding strings and T-duality
of KK and H-monopoles}, hep-th/9708086.
\item E. Eyras, B. Janssen and Y. Lozano, {\it 5-branes, KK monopoles and
T-duality}, hep-th/9806169.
\item A. Font, L. Ibanez, D. Lust and F. Quevedo, \pl 249 (1990) 35; S. J.
Rey, Phys. Rev. D43 (1991) 526.
\item C. Hull and P. Townsend, \np 438 (1995) 109.
\item A. Sen, Int. Jour. Mod. Phys. A9 (1994) 3707.
\item A. Sen, {\it Kaluza-Klein dyons in string theory}, hep-th/9705212.
\item J. Blum, {\it H-dyons and S-duality}, hep-th/9702084.
\item M. Cvetic and D. Youm, \np 438 (1995) 182; \np 449 (1995) 146 (A).
\item M. Cvetic and D. Youm, Phys. Rev. D53 (1996) 584.
\item J. Schwarz and A. Sen, \pl 312 (1993) 105.
\item G. Gibbons and K. Maeda, \np 298 (1988) 741.
\item G. Horowitz and A. Strominger, \np 360 (1991) 197.
\item M. Duff, R. Khuri and J. Lu, Phys. Rep. C259 (1995) 213.
\item A. Das and S. Roy, \np 482 (1996) 119.
\item J. Maharana and J. Schwarz, \np 390 (1993) 3.
\item A. Shapere, S. Trivedi and F. Wilczek, Mod. Phys. Lett. A6 (1991) 2677.
\item M. de Roo, \np 255 (1985) 515.
\item E. Cremmer, B. Julia, H. Lu and C. Pope, {\it Dualisation of dualities
I}, hep-th/9710119.
\item J. Schwarz and A. Sen, \np 411 (1994) 35.
\item J. Schwarz, \pl 360 (1995) 13.
\item S. Roy, \pl 421 (1998) 176.
\item J. X. Lu, \pl 313 (1993) 29.
\item P. Dirac, Proc. R. Soc. A133 (1931) 60; J. Schwinger, Phys. Rev.
144 (1966) 1087; 173 (1968) 1536; D. Zwanziger, Phys. Rev. 176 (1968)
1480, 1489; E. Witten, Phys. Lett. B86 (1979) 283.
\item J. Schwarz, {\it Lectures on superstring and M-theory dualities},
hep-th/9607201.

\end{enumerate}

\vfil\eject

\end{document}